\documentclass[journal]{IEEEtran}
\IEEEoverridecommandlockouts
\usepackage{enumitem}
\usepackage{cite}
\usepackage{amsmath,amssymb,amsfonts}
\usepackage{algorithmic}
\usepackage{graphicx}
\usepackage{textcomp}
\usepackage{xcolor}
\def\BibTeX{{\rm B\kern-.05em{\sc i\kern-.025em b}\kern-.08em
    T\kern-.1667em\lower.7ex\hbox{E}\kern-.125emX}}
\usepackage{amsfonts}
\usepackage{amsmath}
\usepackage{amsthm}
\usepackage{comment}
\usepackage{float}
\usepackage{graphicx}
\usepackage{epstopdf}
\usepackage{multirow}
\usepackage{caption}
\usepackage{subcaption}
\usepackage{ragged2e}
\usepackage{comment}
\usepackage{enumitem}
\usepackage[colorlinks =True,
            linkcolor  =blue,
            anchorcolor=green,
            citecolor  =blue]{hyperref} 
\usepackage[ruled,vlined]{algorithm2e}

\theoremstyle{plain}

\newtheorem*{theorem*}{Theorem}

\def\V{{\mathbf  V}}

\def\bu{\mathbf{u}}
\def\bv{\mathbf{v}} 
\def\bw{\mathbf{w}}

\def\bx{\mathbf{x}} 
\def\by{\mathbf{y}}

\ifCLASSINFOpdf
\else
\fi

\usepackage[ labelsep=period]{caption}
\begin{document}
%
\title{Fast Adaptation for Deep Learning-based Wireless Communications}
%
%
%

\author{Ouya Wang,
        Hengtao He,
        Shenglong Zhou,
        Zhi Ding,
        Shi Jin,
        Khaled B. Letaief,
        and Geoffrey Ye Li
\IEEEcompsocitemizethanks{
\IEEEcompsocthanksitem O. Wang and G. Li are with the ITP Lab, Department of Electrical and Electronic Engineering, Imperial College London, the United Kingdom ( ouya.wang20@imperial.ac.uk, geoffrey.li@imperial.ac.uk).\protect
\IEEEcompsocthanksitem S. Zhou is with the School of Mathematics and Statistics, Beijing Jiaotong University, Beijing, China (shlzhou@bjtu.ed.cn).\protect
\IEEEcompsocthanksitem H. He and K. Letaief are with the Department of Electronic and Computer Engineering, The Hong Kong University of Science and Technology, Hong Kong (eehthe@ust.hk; eekhaled@ust.hk).\protect
\IEEEcompsocthanksitem Z. Ding is with the Department of Electrical and Computer Engineering, University of California at Davis, USA (zding@ucdavis.edu).\protect
\IEEEcompsocthanksitem S. Jin is with the National Mobile Communications Research Laboratory, Southeast University, Nanjing 210096 (jinshi@seu.edu.cn).}

}

\maketitle

\begin{abstract}
The integration with artificial intelligence (AI) is recognized as one of the six usage scenarios in next-generation wireless communications. However, several critical challenges hinder the widespread application of deep learning (DL) techniques in wireless communications. In particular, existing DL-based wireless communications struggle to adapt to the rapidly changing wireless environments. In this paper, we discuss fast adaptation for DL-based wireless communications by using few-shot learning (FSL) techniques. We first identify the differences between fast adaptation in wireless communications and traditional AI tasks by highlighting two distinct FSL design requirements for wireless communications. To establish a wide perspective, we present a comprehensive review of the existing FSL techniques in wireless communications that satisfy these two design requirements. In particular, we emphasize the importance of applying domain knowledge in achieving fast adaptation. We specifically focus on multiuser multiple-input multiple-output (MU-MIMO) precoding as an examples to demonstrate the advantages of the FSL to achieve fast adaptation in wireless communications. Finally, we highlight several open research issues for achieving broadscope future deployment of fast adaptive DL in wireless communication applications.

\end{abstract}


%

\section{Introduction}\label{sec::intro}
%
%
%
%
\IEEEPARstart{T}{he} fifth generation of mobile communications has ushered in a wide range of new services. However, emerging applications, such as the Internet of Everything, Tactile Internet, seamless virtual, and augmented reality continue to demand faster, more reliable, and more efficient wireless communication networks. Artificial intelligence (AI) has triggered technological advancements and innovations across diverse domains. In fact, AI integration with communications has been identified as one of the six usage scenarios in next-generation communication systems (6G), which has led to tremendous growth of interest in developing novel AI technologies include deep learning (DL)-based wireless communications \cite{qin2019deep}. 

DL in wireless communications has grown in popularity in recent years. Indeed, DL has demonstrated its potential for enhancing performance of existing algorithms, achieving near-optimal results, enabling end-to-end system optimization, and supporting entirely novel use cases that are not feasible with traditional model-based approaches. However, there still exist several critical challenges that hamper the widespread adoption of DL in wireless networks. First, wireless channels are changing dynamically due to temporal environment variations, which degrades the performance of the trained DL-based system. Therefore, it is critical to continuously re-collect sufficient data and retrain the model whenever the wireless environment changes, which is impractical as data collection and model retraining are time-consuming. Furthermore, the limited computational resources at mobile devices presents another challenge as DL fundamentally depends on highly parameterized architectures and requires powerful devices for large-scale training. Therefore, DL-based wireless communication systems must quickly adapt to dynamic environments with a minimal computational overhead. For instance, channel estimation in time-varying environments requires DL-based systems to adapt rapidly changing to new channel conditions as channel characteristics only remain static for short durations.

One solution to fast adaptation in response to new environments/tasks utilizes few-shot learning (FSL) techniques, in traditional AI frameworks. FSL techniques in traditional AI tasks rely on learning prior experience obtained from the previous tasks to compensate for the lack of training data in new tasks. However, applying the FSL techniques for wireless communications directly is challenging. These FSL techniques require substantial training resources to obtain effective experience corresponding to new tasks or new environments, which is impractical for wireless communications due to limited computational resources and storage capacity at user devices. 

To develop effective FSL techniques applicable to wireless communication, it is essential to take the features of the wireless tasks into consideration. Fortunately, some traditional data-driven FSL techniques can be tailored for wireless tasks. Furthermore, domain knowledge in wireless communications enables to integrate prior experience into DL-based systems, rather than relying exclusively on data-driven learning. Therefore, both data-driven FSL techniques and domain knowledge can be utilized to achieve fast adaptation for DL-based wireless communications. In this article, we provide a comprehensive overview on fast adaptation for wireless communications achieved by data-driven FSL techniques and domain knowledge. We will emphasize the importance of domain knowledge, especially model-driven DL approaches, and highlight promising directions for future research. 


The rest of this article is organized as follows. In Section \ref{sec:problem_settings}, we introduce mechanisms of FSL techniques in conventional AI tasks, explain the challenges on applying them to wireless communications, and highlight some features of wireless tasks that can be potentially used for fast adaptation. Section \ref{section::algorithm} elaborates the data-driven FSL techniques for fast adaptation in wireless communications. Mechanisms of how wireless domain knowledge can reduce the demand on data and enable fast adaptation in DL-based wireless systems are introduced in Section \ref{section::model}


\section{FSL in AI and Wireless Communications} \label{sec:problem_settings}

In this section, we review FSL techniques in traditional AI tasks, elaborate the challenges of applying existing FSL techniques to wireless tasks, and then outline the requirements for achieving fast adaptation in wireless communications.

\subsection{FSL Techniques for Traditional AI Tasks} \label{sec::fsl_ai}
FSL techniques aim to facilitate DL models generalize and perform well on new tasks with only few training examples. The primary challenge in FSL is that deep neural networks (DNNs) trained on small datasets often fail to generalize effectively. Due to limited data, the approximation of DNN is unrealiable and prone to overfitting \cite{wang2020generalizing}. To solve this issue, prior knowledge must be leveraged to enhance approximation ability of DNN. This enhancement can be achieved through \textbf{data processing} to enlarge the training dataset; \textbf{structure-based approaches} to constrain the hypothesis space; and \textbf{learning algorithms} to optimize search strategies \cite{wang2020generalizing}.
\begin{itemize}[leftmargin=10pt]
\item \textbf{Data processing} augments training samples to enrich supervised information for new tasks. A prevalent augmentation strategy involves generating multiple variants of each training sample through geometric transformations. This transformation can be learned from any similar class by iteratively aligning each sample with other samples \cite{wang2020generalizing}. 
\item \textbf{Structure-based approaches}, such as embedding learning, map the high-dimensional samples into lower-dimensional vectors to facilitate the differentiation of samples from different classes. Smaller hypothesis space can be constructed for lower dimensional data, thus requiring fewer training samples. Embedding learning is primarily designed for classification tasks as it focuses on embedding data into a feature space that maximizes class separation. 
\item \textbf{Learning algorithms} consist of two main methods that can optimize training strategies and tackle FSL problems. First, the initialization of DNN can be optimized, to the extent only a few fine-tuning iterations are needed to train the model, enabling the model to perform well in new tasks with limited samples. Second, an additional model is trained to guide each optimization step by dynamically adjusting hyper-parameters, e.g. step size and learning rate, thus accelerating the training process \cite{wang2020generalizing}.
\end{itemize}

\begin{figure}[h]
\centering
\includegraphics[scale=.345]{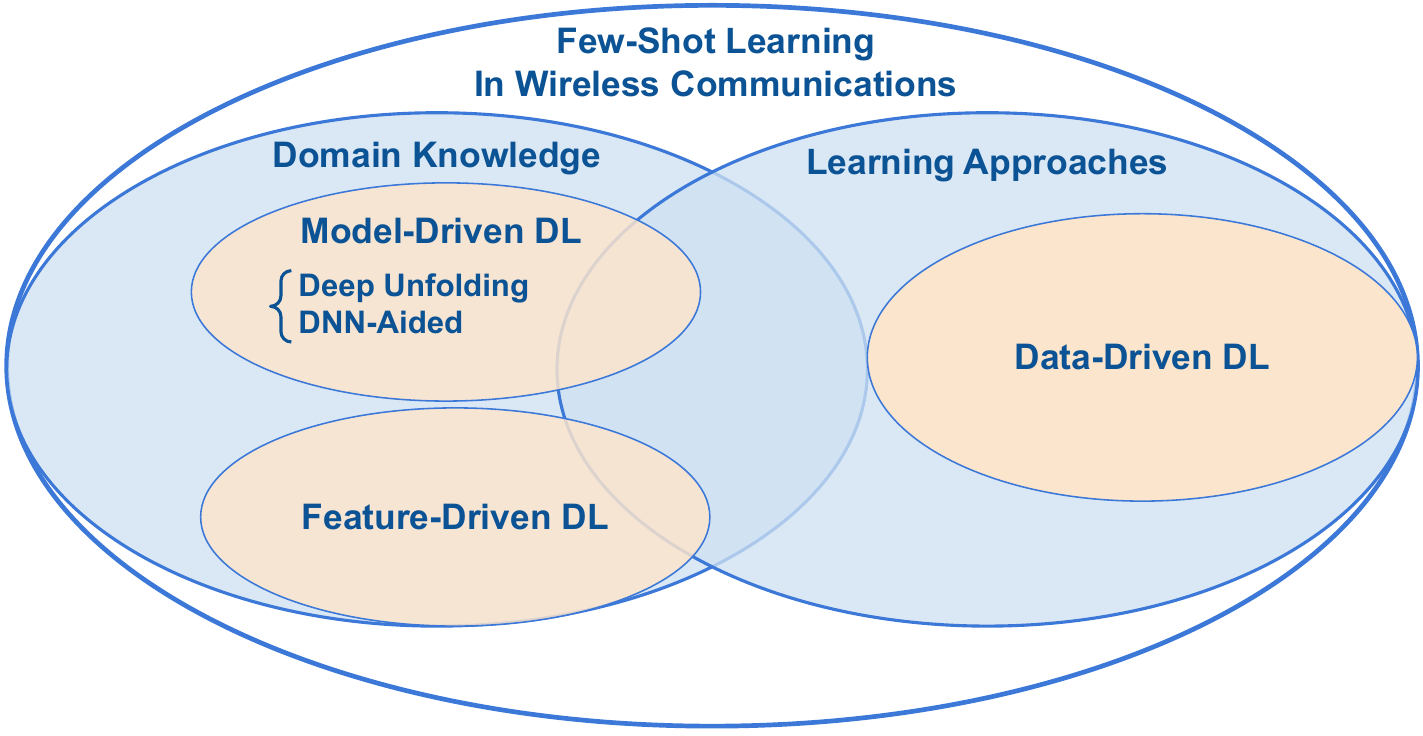}
\caption{Categories of FSL techniques in wireless communications.}\label{fig:concept}
\end{figure}
\subsection{Challenges of FSL in Wireless Communications}\label{sec::challenges}
The obstacles to employ conventional FSL techniques to wireless communications can be elaborated from two perspectives, task settings and practical considerations. We define different wireless tasks based on varying characteristics of the wireless environments, such as power delay profiles and signal-to-noise ratios (SNRs). Many FSL tasks in AI have different \textbf{task settings} from those in wireless tasks. Typically, AI tasks are oriented towards classification. The objective of corresponding approaches, such as structure-based approaches, is to assign discrete labels to static input features. However, wireless tasks such as signal detection, channel estimation, and resource allocation, often involve continuous regression and real-time decision-making in dynamic environments. These tasks extend beyond the discrete and static nature of classification problems.

The \textbf{practical consideration} of wireless systems also leads to the application challenges. For instance, one category of FSL techniques for AI tasks involves aligning the feature distributions of new tasks with those of training tasks. This alignment enables to address new tasks in a similar domain, using models that require minimal fine-tuning and significantly less data. However, these approaches require training tasks data in the adaptation stage, necessitating numerous training iterations. In wireless communications, accessing data from previous tasks in the adaptation stage is not feasible. Additionally, feature alignment requires a substantial number of training iterations. On user devices or terminals, this process is hampered by limited processing power and fails to achieve fast adaptation.

\subsection{Characteristics and Requirements of FSL in Wireless Communications }\label{sec::conditions}
To address the challenges outlined in Section \ref{sec::challenges}, it is essential to exploit the unique characteristics of FSL tasks in wireless communication. There are three vital features: 
\begin{itemize}[leftmargin=10pt]
\item Adaptation efficiency should be considered because of the dynamic nature of wireless environments. A slow adaptation process may lead to significant performance degradation because the wireless environment could have already changed before the system finishes adapting. 
\item User devices are constrained by limited processing power and hardware capabilities. i.e., due to limited storage, unable to access data from previous tasks. 
\item Wireless domain knowledge can be employed to design FSL techniques. In wireless communications, domain knowledge, including models, algorithms, and unique attributes developed through extensive research, serves as a form of prior experience. Unlike traditional AI methods that rely solely on data-driven approaches, wireless domain knowledge can enhance problem-solving capabilities in DL-based wireless tasks.
\end{itemize}

We focus on the three characteristics of FSL tasks in wireless communications, the corresponding requirements: limited training cost and accessing only data from the new task. The training cost, including the number of parameters that need to be updated, the amount of data required for training, and the number of training iterations, should be limited to facilitate the system adapt to new tasks quickly. Only data in new tasks can be accessed when processing new tasks, without using any data from previous tasks. In addition to strictly adhering to these two requirements, it is essential to incorporate domain knowledge when developing FSL techniques, rather than fully relying on data-driven methods.


The exiting FSL techniques satisfying the mentioned requirements can be classified into two categories, as demonstrated in Fig. \ref{fig:concept}. The first one is that DL-based wireless systems are developed in purely data-driven manners, without incorporating any wireless domain knowledge. In this case, we can optimize FSL techniques designed for AI tasks to satisfy the two design requirements illustrated above. These techniques can then be employed to achieve fast adaptation in wireless communications through data-driven manners. The second category focuses on incorporating domain knowledge for DL-based wireless systems. For instance, by integrating classical algorithms, the complexity of parameterizing these systems can be significantly reduced. This approach constrains the hypothesis space of the DL-based system using domain knowledge, thereby reducing the data demand for adapting to new tasks. In Sections \ref{section::algorithm} and \ref{section::model}, we will present several applications of FSL techniques for wireless communications in detail.

\section{Data-Driven FSL for Wireless Communications}\label{section::algorithm}

Data-driven FSL techniques include data processing, structure-based approaches, and learning algorithms as indicated in Section \ref{sec::fsl_ai}. However, neither structure-based approaches nor data processing approaches can achieve fast adaptation in wireless communications. This is because structure-based approaches are primarily designed for classification tasks, and are not suitable for tackling wireless tasks due to different task settings. Furthermore, data processing approaches address data scarcity by developing self-supervised learning techniques \cite{shlezinger2020viterbinet} or wireless-oriented data augmentation strategies \cite{raviv2023data} to generate more labelled data, which provide sufficient data for new tasks adaptation but increase the training cost. This issue is particularly pronounced for mobile devices, as longer processing time is required due to their limited processing power, leading to less efficient adaptation.

\begin{figure}[h]
\centering
\includegraphics[scale=.340]{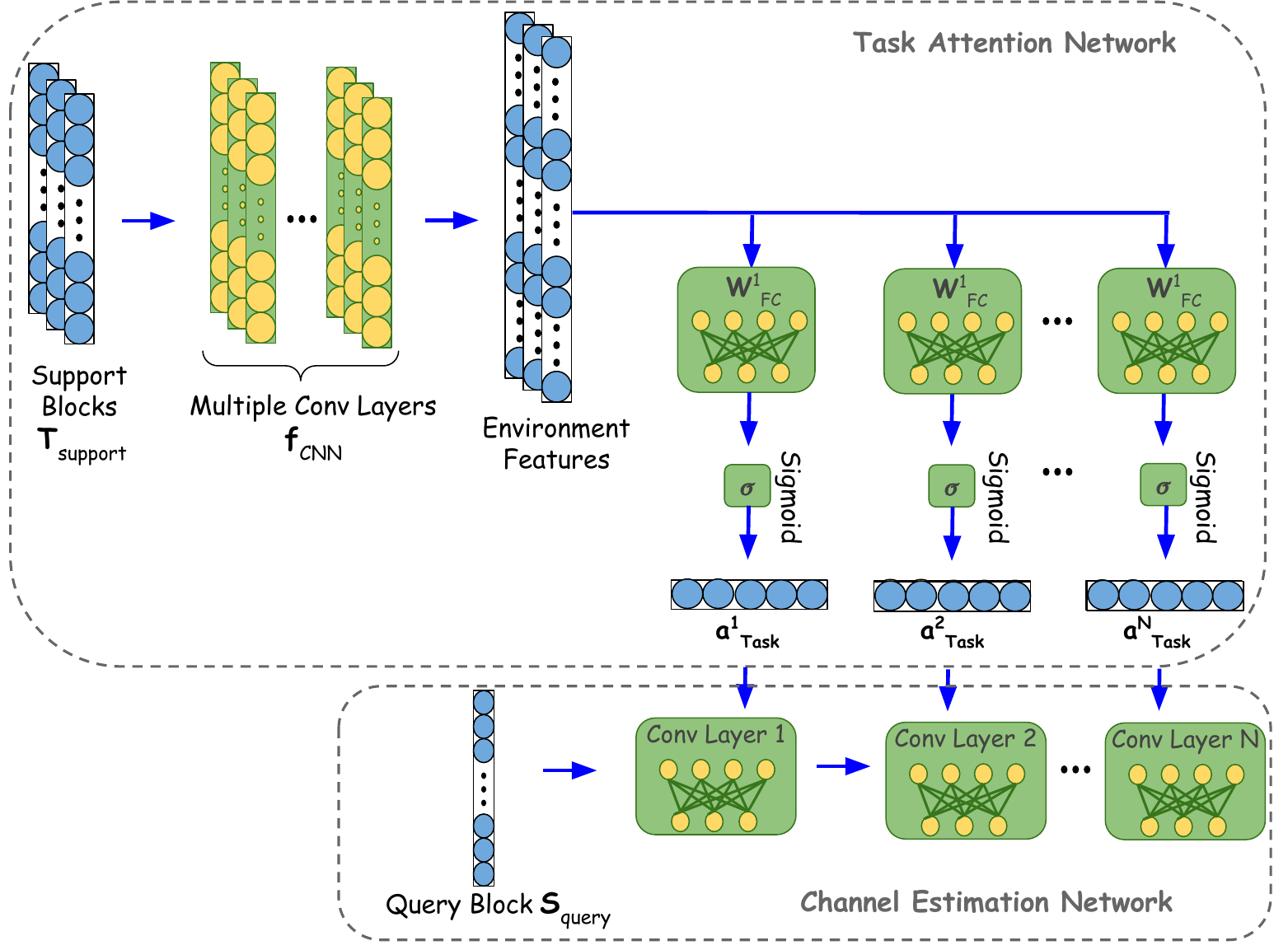}
\caption{Structure of hypernetwork to generate parameters for channel estimation network.}\label{fig:Task_attention}
\end{figure}

Given the difficulties associated with data processing and structure-based FSL approaches in achieving fast adaptation for wireless communications, learning algorithms have emerged as only promising candidates in data-driven FSL techniques. Learning algorithms utilize the prior experiences from previous tasks to refine strategies for new tasks, with limited training samples and minimal training cost. This section introduces the application of two learning algorithms, meta learning and multi-task learning, for wireless communications.

\subsection{Meta Learning} 

Meta learning uses data from multiple tasks to optimize the training process, enabling efficient adaptation to the prior unknown learning task using only few samples \cite{wang2020generalizing}. This approach treats the training process as an accumulation of learning experience from known tasks. Depending on strategies of learning experience, meta learning in wireless communications can be divided into model-based and optimization-based meta learning.

\begin{itemize}[leftmargin=10pt]
 \item \textit{Model-based meta learning} employs an additional model, e.g. a hypernetwork, to learn the experience from the previous tasks and apply it to new tasks. The hypernetwork utilizes limited training data for each task to generate task-specific weights, dynamically adjusting the DL system to generalize well across various tasks while adapting to new tasks. Model-based meta learning has been successfully applied in DL-based OFDM channel estimation \cite{wang2022learn}. This application is developed under the scenario where only few subcarriers contain pilots in OFDM pilot blocks. In this case, traditional channel estimation algorithms, such as least square (LS) and minimum mean-squared error (MMSE), cannot work well. Therefore, the convolutional neural network (CNN) has been utilized for data-driven channel estimation. As illustrated in Fig. \ref{fig:Task_attention}, the CNN estimates the channel coefficients from pilot blocks, referred to as query blocks. The few-shot training samples, utilized as support blocks, provide supervised information for each task. In the conventional AI tasks, few-shot training samples leverage the true values/labels to provide supervised information for adaptation. However, obtaining the true CSI for support blocks in real-world scenarios is extremely difficult. To address this challenge, we increase the number of pilots in each support block. More accurate CSI can be obtained from the support blocks compared with the query blocks, serving as supervised information in channel estimation.
 
 The hypernetwork in Fig. \ref{fig:Task_attention}, also known as task attention network, first learns to extract statistic environment features, such as power delay profile, using few support blocks. These environmental features are then employed to generate weight vectors for the output feature maps in each layer of the channel estimation network. The experiment initially evaluates how the number of support blocks affects adaptation performance, ultimately selecting 16 support blocks for further testing. Subsequent adaptation experiments in four new environments demonstrate that the proposed approach outperforms state-of-the-art (SOTA) DL-based methods.
 

 \item \textit{Optimization-based meta learning} shares the experience across tasks through optimizing the hyperparameters of DL-based system, e.g. initialization. A notable example of initialization-based scheme is the model-agnostic meta learning (MAML) algorithm \cite{finn2017model}, which finds an initialization that allows quick adaptation to new tasks with few training samples. MAML has been utilized to jointly design an encoder and decoder that can quickly adapt to new channel coefficients, as demonstrated in \cite{park2020learning}. Both the encoder and decoder are implemented using neural networks. A message is first encoded into a codeword at the transmitter, and the received signal, determined by channel coefficients, is decoded into an estimated message. The primary objective is to accurately reproduce the input message at the output of the decoder. However,optimizing these networks typically requires many training iterations to minimize block-error rates for new channel conditions. The MAML algorithm addresses this issue by learning an initialization close to optimal for various tasks. With this optimal initialization, only one-step training iteration is required for the system to achieve lower block-error rate compared with other learning-based approaches, such as joint learning.

\end{itemize}

\subsection{Multi-Task Learning} 

Multi-task learning enables DL-based systems to address multiple tasks simultaneously by structuring the system into shared and task-specific parts. The shared part remains consistent across all tasks, designed to learn task-agnostic features among them. In contrast, the task-specific part is uniquely tailored to each individual task, focusing on learning task-specific features. In new tasks adaptation, only the task-specific part requires fine-tuning, which offers two advantages. First, the data demand for new task adaptation is reduced with fewer training parameters. Second, the generalization performance to new tasks is enhanced due to task-agnostic features. These features facilitate effective generalization across various tasks, including new tasks, and help prevent the entire system from overfitting to few-shot samples.

\begin{figure}[h]
\centering
\includegraphics[scale=.345]{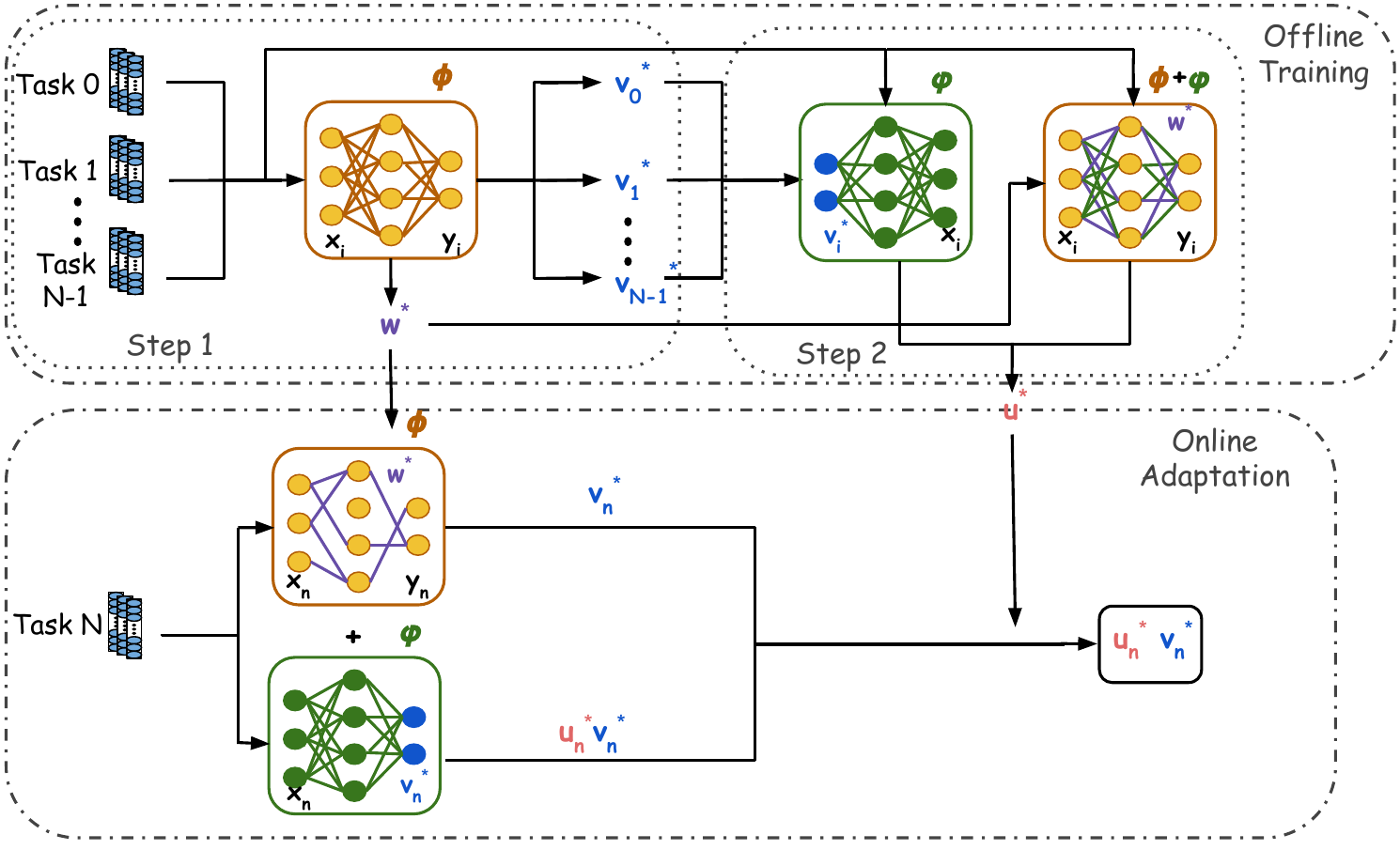}
\caption{Structure of multi-task meta learning. $\phi$ and $\varphi$ refer to the DL-based system and the hypernetwork.}\label{fig:OA}
\end{figure}

Multi-task learning has been successfully applied to fast adaptation in mmWave beamforming \cite{wang2023new}. The entire learning process is demonstrated in Fig. \ref{fig:OA}, where ${\bw}^*$ is denoted as the shared part and ${\V^*:=( {\bv}_0^*,{\bv}_1^*,\cdots, {\bv}_{N-1}^*)}$ is denoted as $N$ task-specific parts for $N$ tasks. Step 1 aims to learn ${\bw}^*$ and ${\V^*}$ by solving the following optimization problem:  

\begin{eqnarray}\label{phase-i-opt}
\begin{array}{ll}
({\bw}^*,&{\bv}_1^*,\cdots,{\bv}_N^*)\\ 
&:=\underset{\bw,\bv_1,\cdots,\bv_N}{\rm argmin} ~\frac{1}{N} \sum_{n=1}^N \sum_{i=1}^{d_n} \ell\left(f(\bw,\bv_n;\bx^i_n),\by^i_n \right), \nonumber
\end{array}
\end{eqnarray}
where $\ell$ is the loss function, $d_n$ is the number of training samples in each task and $f(\cdot,\cdot;\bx)$ is the DL system with input channel $\bx$ estimated from pilots, and $\by$ is the corresponding beam vectors to maximize the data rate. Compared with directly fine-tuning the task-specific part for new tasks, we can learn a mapping from input channel $\bx_n$ to its corresponding $\bv_n$ for each task $n$ and use this mapping to predict the task-specific part for new tasks. As shown in Fig. \ref{fig:OA}, a hypernetwork parameterized by $\bu^*$ in step 2 is employed to learn this mapping from training tasks, thereby obtaining experience in predicting task-specific parts using input channel across $N$ tasks. To further enhance adaptation efficiency, MAML is utilized to optimize the initialization of the hypernetwork. This allows the hypernetwork to be fine-tuned in only ten training iterations, enabling faster acquisition of task-specific part ${\bv}_n^*$ after applying MAML. Experimental results in \cite{wang2023new} demonstrate that the combination of multi-task learning and meta learning achieves better adaptation performance compared to the existing learning-based approaches, such as transfer learning and meta-transfer learning while also maintaining high adaptation efficiency.


Approaches proposed in \cite{wang2022learn} and \cite{wang2023new} both learn shared patterns from multiple training tasks and employ hypernetworks to generate task-specific weights. The approach in \cite{wang2023new} introduces constraints and penalty terms to mitigate overfitting during adaptation to new tasks. Furthermore, it can clearly identify the shared and task-specific features through the formulation of the optimization problem. This formulation enables a more accurate characterization of task features than the method in \cite{wang2022learn}, which tends to blur distinctions due to joint training. However, the approach in \cite{wang2022learn} is more efficient during adaptation, as it only requires fine-tuning the channel estimation network with only five convolutional layers, whereas \cite{wang2023new} requires fine-tuning both the hypernetwork and main system, leading to higher training costs.


\begin{figure*}[h]
\centering
\includegraphics[scale=.455]{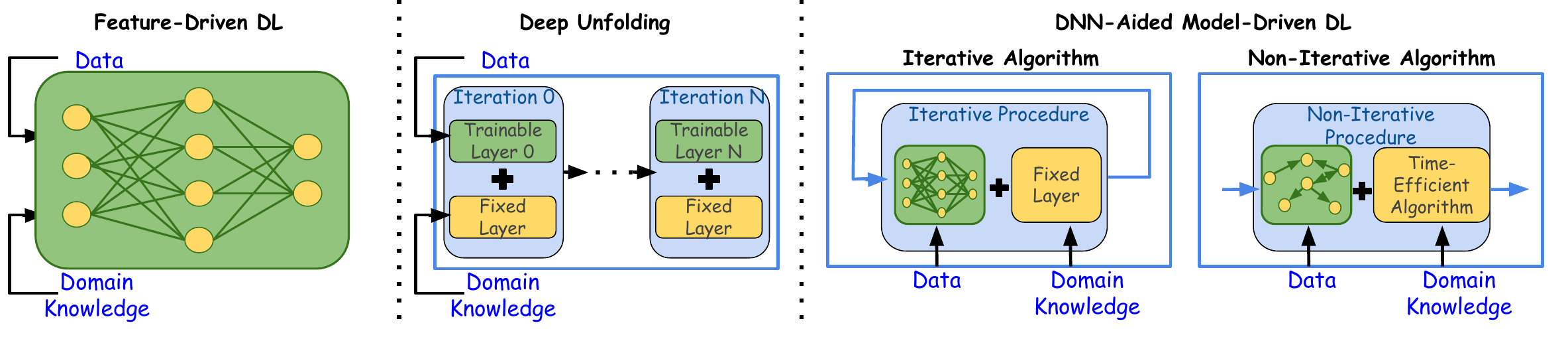}
\caption{The structures of feature-driven DL, deep unfolding, and DNN-aided model-driven DL designed for both iteration or non-iteration algorithms. The FPN is developed for iterative algorithms, and the NC framwork is designed for non-iterative algorithms.}\label{fig:blackbox}
\end{figure*}
\section{Domain Knowledge-based FSL Techniques}\label{section::model}

DL-based wireless system can be developed by incorporating domain knowledge \cite{he2020model}, rather than relying solely on the huge volume of data to configure neural networks (NNs). This approach, known as non-blackbox DL, combines domain knowledge and DL to construct the network. Non-blackbox DL can be further divided into two categories: feature-driven DL and model-driven DL. Feature-driven DL utilizes distinctive wireless characteristics for system configuration without employing predefined models or algorithms. In contrast, model-driven DL integrates classic wireless models and algorithms into DL system design.

\subsection{Feature-Driven DL}


Feature-driven DL exploits domain knowledge to preprocess the input data before feeding it into the DL-based system, thus improving system performance. It can achieve fast adaptations as well. In particular, the plug-in CSI feedback scheme, proposed in \cite{liu2024deep}, is equipped with a lightweight translation module designed to fit dynamic wireless environments. This CSI feedback system is initially pretrained on training tasks. The lightweight translation module processes CSI data from new tasks, aligning it with the data format in training tasks, which ensures efficient reuse of the pretrained system. Therefore, for new tasks, only the translation module need fine-tuning. Originally developed for image tasks using data-driven methods, conventional translation modules are often highly parameterized. In contrast, leveraging domain knowledge allows for a reduction in module parameters. Specifically, the inherent sparsity of the CSI matrix in the angular delay domain is considered as domain knowledge and facilitates the utilization of the circular shift to promote the similarity in data distribution between new and training tasks. This preprocessing step is applied to CSI data before sending it to the translation module. By incorporating this step, a lightweight translation module can be developed, enabling quick adaptation to new tasks with minimal CSI data.

\subsection{Model-Driven DL}

Model-driven DL utilizes classic models and algorithms as domain knowledge to develop DL-based systems. In particular, the mathematical algorithm is one of the important parts for domain knowledge, such as, weighted minimum mean-squared error (WMMSE) is the classic precoding for multi-user interference suppression and sum-rate maximization \cite{hu2020iterative}. These algorithms depend heavily on accurate modeling to effectively address wireless problems and inaccuracy in modeling leads to serious performance degradation. Furthermore, some of them have high computational complexity, which limits their practical implementation. Model-driven DL constructs the network architecture using the model and the algorithm, where DL can compensate for performance loss resulting from inaccurate modeling \cite{he2019model}. Additionally, the computational complexity can be reduced by replacing high-complexity operations with DL.

Model-driven DL inherits advantages of classic algorithms, which allows DL systems to benefit from domain knowledge by reducing the number of trainable parameters. Different from feature-driven DL, model-driven DL primarily constructs upon existing algorithms and employs DL as auxiliary steps to address specific challenges. It helps prevent overfitting in DL-based systems trained with limited data. However, if the underlying algorithm is not well-suited for certain problem scenarios, model-driven DL may not perform effectively. In such cases, the system can be developed using feature-driven DL to incorporate specific characteristics of wireless communications or by employing a purely data-driven approach. Model-driven DL can be further divided into two categories: deep unfolding and DNN-aided model driven DL.

\subsubsection{Deep Unfolding}
As indicated by its name, this approach unfolds a traditional iterative algorithm into a neural network by introducing trainable parameters at each iteration \cite{he2020model}, as illustrated in Fig. \ref{fig:blackbox}. For example, a multiple-input multiple-output (MIMO) detector is established by unfolding the expectation propagation (EP) algorithm, with several trainable parameters \cite{zhang2020meta}. In particular, each iteration may have only one trainable parameter, significantly reducing the total number of parameters, which decreases computational requirements in the fast adaptation stage. To further improve the adaptation efficiency, a structure-based meta learning algorithm uses a recurrent neural network (RNN) to update trainable parameters, thereby achieving faster convergence when adapting to new tasks.



\subsubsection{DNN-aided model-driven DL}
The other category of model-driven DL integrates the DNN into the framework of classic algorithms, which is referred to as DNN-aided model-driven DL. These classic algorithms can be either iterative algorithms or non-iterative algorithms. The structure of DNN-aided model-driven DL is demonstrated in Fig. \ref{fig:blackbox}

The neural calibration (NC) framework is proposed to improve the traditional \textbf{non-iterative algorithms} \cite{ma2022learn}, by employing a structured NN to calibrate the input data of low-complexity algorithms. The NC framework is compatible with various transceiver modules by utilizing different differentiable algorithms as the basis of calibration. For example, zero-forcing (ZF) or maximal ratio transmission (MRT) can be selected as basic algorithms for near-field beam focusing tasks \cite{ma2022learn}. 

A unique architecture is further developed for the NN in the NC framework. This unique design ensures that the NN has the permutation equivariance (PE) property and naturally inherits the structure of wireless network topology. The PE property is obtained by ignoring irrelevant indexing issues during NN training. It reduces the data demand since the NN does not need to learn robust features by incorporating every possible permutation of the input. Furthermore, the NN's ability to generalize across permutations demonstrates a solid foundation of learned representations applicable to various input scenarios, including new tasks, thereby reducing the amount of data required for adaptation.

An \textbf{iterative algorithm} after convergence can be interpreted as the equilibrium state achieved through infinite fixed point iterations. Based on this assumption, a fixed point network (FPN), proposed in \cite{yu2023adaptive}, is introduced as a versatile framework to construct iterative algorithms by utilizing the fixed point iteration with learnable contraction mapping. FPN has been proved to be with linear convergence inheriting from fixed point iteration. It maintains linear steps of iterative algorithms while replaces non-linear steps by NNs to enhance performance. Specifically, a NN is employed to learn mapping from fixed point to the desired optimum solution, which aims to preserve the merits of classical algorithms largely.


FPN has been applied to achieve fast adaptation in channel estimation. The system is first pre-trained offline and then directly deployed in the test stage. If the potential performance drop is detected due to channel distribution shifts, the parameters of the pre-trained model will be fine-tuned using one particular received pilot signal. According to the experiment in \cite{yu2023adaptive}, the overhead of one iteration fine-tuning is comparable to a single forward propagation. Typically, only five fine-tuning iterations are sufficient. The fast convergence of FPNs is attributed to the linear convergence propoerty of fixed-point iterations. Therefore, FPNs can adapt to new tasks with an accelerated convergence speed.

\section{Numerial Results}
We take the multiuser MIMO (MU-MIMO) precoding design as an example to compare the performance of deep unfolding, FPN, and MAML. Deep unfolding, as inspired by \cite{hu2020iterative}, replaces matrix inversion with linear operations involving trainable variables, thereby reducing computational complexity. The FPN substitutes WMMSE's nonlinear operations with learning-based estimators while maintaining linear operations. To demonstrate the advantages of domain knowledge, both methods are trained jointly on all tasks and fine-tuned using few-shot samples in the new task. Furthermore, MAML is integrated with these model-driven DL approaches. To further highlight the benefits of using MAML, it is incorporated with a blackbox DL, which is developed in a data driven manner without domain knowledge and directly generates precoding matrices from CSI.


Our experiment considers the MU-MIMO system with four users, eight transmitter antennas, and two receiver antennas operating over a narrow-band channel. We utilize the correlated channel model to simulate various channel distributions by adjusting correlation coefficients for channel correlation matrices. We generate eight training tasks and one testing task, with 16 CSI for each new task adaptations. The system is initially trained under the SNR$=$20dB scenario but evaluated under different SNR scenarios.

\begin{table}[t]
\renewcommand{\arraystretch}{1.5}\addtolength{\tabcolsep}{-3.5pt}
    \centering
    \caption{The system testing performance which is trained in each training task and directly deployed into the testing task (SNR = 20dB)}
    \label{table:1}
    \begin{tabular}{cccccccccc}
        \hline
         Training  & Task 1 & Task 2 & Task 3 & Task 4 & Task 5 & Task 6 & Task 7 &Task 8\\
       \hline
       \multicolumn{9}{c}{Sum-Rate of the Testing Task (bits/s/Hz)}\\\hline
      Unfolding & 12.76 & 12.58 & 12.54 & 12.16 & 11.79 & 11.23 & 11.24 & 9.86 \\ 
       FPN & \textbf{12.86} & \textbf{12.71} & \textbf{12.68} & \textbf{12.31} & \textbf{12.04} & \textbf{11.92} & \textbf{11.52} & 10.44 \\ 
       Blackbox & 12.08 & 11.65 & 11.57 & 11.67 & 9.92 & 10.28 & 10.81 & \textbf{11.18} \\ 
       \hline

    \end{tabular}%
    
    \medbreak
    \end{table}

\begin{figure}[th]
\centering
\includegraphics[width=0.48\textwidth]{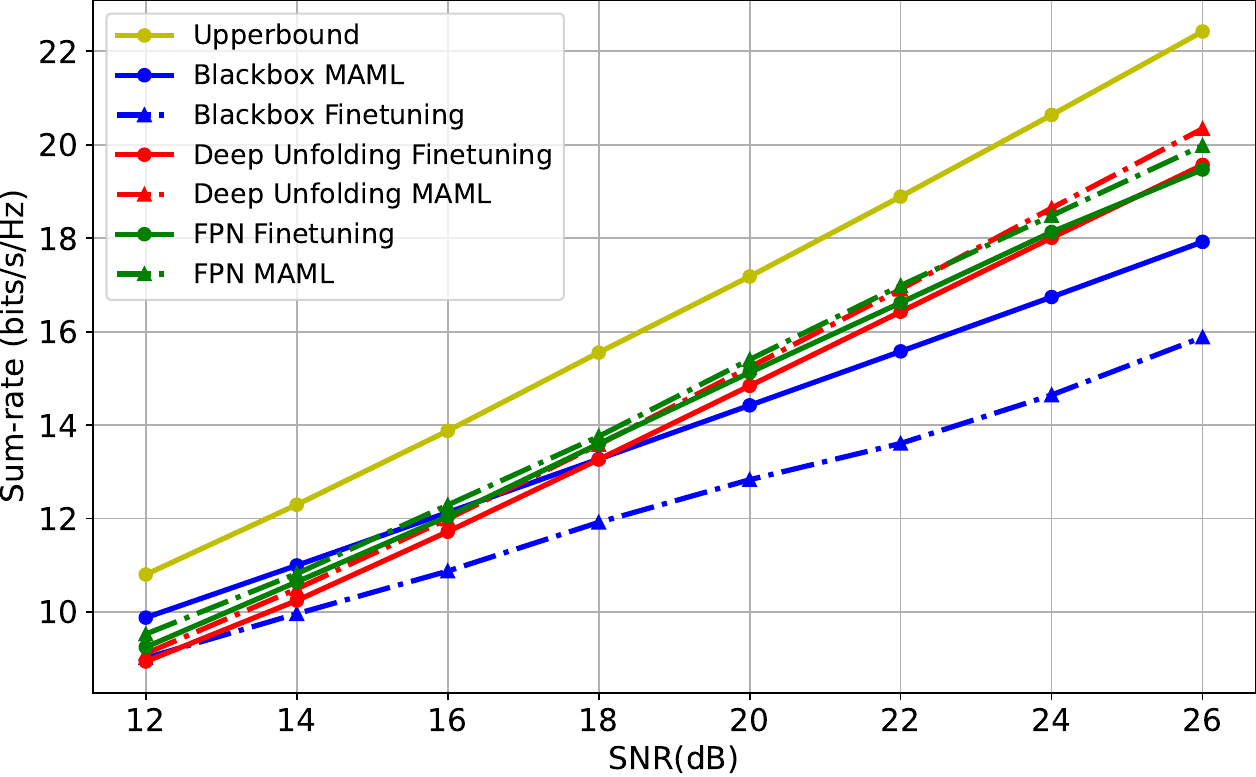}
\caption{The fast adaptation performance of new tasks, using learning algorithms and domain knowledge. The experiment "Upperbound" refers to the scenario where sufficient data is available in the new environment to train the deep unfolding network. }\label{fig:fsl_magazine}
\end{figure}

We will make sure the adaptation performance on the new task is not attributed to high similarity between training and testing tasks. Blackbox DL, deep unfolding, and FPNs are trained on each training task and subsequently evaluated in the testing task, as shown in Table \ref{table:1}. Fig. \ref{fig:fsl_magazine} illustrates the advantages of using domain knowledge and learning algorithms to enhance the adaptation performance of the new task. The testing performance in Table \ref{table:1} is significantly lower than that observed at the SNR$=$20dB in Fig. \ref{fig:fsl_magazine}. Performance degradation caused by overefitting is also illustrated in Fig. \ref{fig:fsl_magazine} for directly fine-tuning highly parameterized blackbox DL models. The blackbox DL performs significantly better when using MAML, requiring 6 fine-tuning iterations for adaptation. Domain knowledge can also enables fast adaptation. Deep unfolding and the FPN requiring 20 and 15 fine-tuning iterations, respectively, to achieve optimal performance. Furthermore, integrating MAML with deep unfolding and the FPN results in substantial improvements in adaptation efficiency, reducing the fine-tuning iterations to approximately five or six. These experiment results confirm that both learning algorithms and domain knowledge are effective for achieving fast adaptation in wireless communications.

\section{Conclusion and Future Research Directions}
Although previous studies have obtained promising results on fast adaptation for wireless communications, this area is still in its early stage. Future research should address open issues such as continual adaptation, few shots accumulation, and large language model (LLMs) for fast adaptation.

\subsection{Continual Adaptations to New Environments}

Currently, our FSL techniques are implemented through one-time adaptation, meaning that communication systems only adapt to a single new environment and then complete the entire adaptation process. This scenario is akin to taking a plane from one location to another, which causes sudden change of the environment. A more practical scenario resembles driving a car, where the environment continuously changes, necessitating the communication system to continually adapt to new tasks. As the system completes adapting to a new environment, information acquired from this environment can be retained as experience to address subsequent environmental changes and facilitate further adaptation.


\subsection{Few Shots Accumulation}
In real-life, wireless data are received sequentially over time. Our current FSL techniques require a fixed number of samples from new tasks, making adaptation less efficient as the system must wait to collect the required number of data samples. Moreover, after adaptation with these few samples, these approaches terminate further optimization even if additional data becomes available later in the same environment. It would be more practical and efficient to develop methods that enable real-time model optimization as data continues to be received in the same new environment. These approaches keep enhancing adaptation performance with ongoing accumulation of wireless data.

\subsection{Large Language Model for Fast Adaptation}
LLM is particularly effective in achieving fast adaptation due to its pre-training on vast and diverse datasets, which endows them with substantial contextual knowledge. Therefore, LLM is able to generalize well to new tasks, which reduces data demand for new task adaptation. LLM offers additional advantages in wireless communications. In particular, the capacity of LLM to learn from extensive datasets helps to manage the complexity and dynamic nature of wireless environments effectively. Moreover, its proficiency in modeling complex patterns and dependencies can enhance the robustness of wireless communication systems, leading to overall better performance.

\bibliographystyle{IEEEtran}
\bibliography{references}

\end{document}